# Boundary element method for normal non-adhesive and adhesive contacts of power-law graded elastic materials

Qiang Li and Valentin L. Popov

Technische Universität Berlin,
Department of System Dynamics and the Physics of Friction,
10623 Berlin, Germany

Corresponding author: qiang.li@tu-berlin.de

**Abstract**

Recently proposed formulation of the Boundary Element Method for adhesive contacts has been generalized for contacts of functionally graded materials with and without adhesion. First, proceeding from the fundamental solution for single force acting on the surface of a half space with a power-law varying elastic modulus, the deformation produced by constant pressure acting on a rectangular element was calculated and the influence matrix was obtained for a rectangular grid. The inverse problem for the calculation of required stress in contact area from a known surface deformation was solved by use of conjugate-gradient technique. For the transformation between the stresses and displacements, the Fast Fourier Transformation is used which drastically reduces the computation time. For the adhesive contact of graded material, the detachment criterion based on the method of Pohrt and Popov was proposed. A number of numerical test for the problem having exact analytical solution have been carried out confirming the correctness of underlying ideas and numerical implementation.

## 1. Introduction

In the last few decades, various methods have been developed for improvement of thermomechanical properties of components, many of them based on using layered structures as coatings, plating techniques or layer lamination process (Bhushan & Gupta, 1991; Roos, Celis, Fransaer & Buelens, 1990; Erdogan, 1995). A logical generalization of this approach is the use of functionally graded materials (FGMs), which became increasingly popular since 1990s. The gradually varying composition and structure of FGMs result in the continuous changes in properties of materials, thus solving some of typical problems of layered materials as poor interface strength and residual stresses. Living species have "discovered" FGMs millions of years ago. Gradient media can be found in many biological structures as skin, bones or bamboo tribes (Jha, Kant & Singh, 2013). Lots of work has been carried out for developing of manufacturing techniques of this kind of materials and for studying their behavior (Birman, Keil & Hosder, 2013; Udupa, Rao & Gangadharan, 2014; Selvadurai, 2007).

Contact problems involving gradient materials were initially studied by researchers in soil mechanics and geomechanics because the variation of elastic properties of soils and rocks strongly affects the settlement and stability of foundations (Borowicka, 1943; Taylor, 1948; Gibson, 1967). In most cases the spatial variation of elastic modulus with depth was assumed to follow either an exponential or a power law; a detailed review can be found in the paper (Selvadurai, 2007). Analytical solutions for the general power law dependence of elastic modulus of an elastic half space was provided for different types of surface loading by Booker et al. (Booker, Balaam & Davis, 1985a; Booker, Balaam & Davis, 1985b), and solutions for both exponential and power-law dependencies were given later by Giannakopoulos et al. (Giannakopoulos & Suresh, 1997a; Giannakopoulos & Suresh, 1997b). For axisymmetric contacts with a power law elastic modulus, all the existing results can be reproduced very easily using the Method of Dimensionality Reduction (MDR) (Heß, 2016).

For contacts of complicated geometries or contacts under arbitrary loading, it is usually impossible to obtain an analytical solution for both elastically homogenous and nonhomogeneous materials, so that the numerical methods are widely developed for contact problems, e.g. finite element method (FEM) (Hyun, Pei, Molinari & Robbins, 2004; Abali, Völlmecke, Woodward, Kashtalyan, Guz & Müller, 2012), boundary element method (BEM) (Putignano, Afferrante, Carbone & Demelio, 2012; Paggi & Ciaveralla, 2010; Pohrt & Popov, 2012; Aleynikov, 2010) and molecular dynamics simulation (Yang & Persson, 2008; Campana & Muser, 2007). For contacts of linearly elastic or viscoelastic materials (Kusche, 2016). The most suitable method for contact problems is the boundary element method as it uses discretization only of the two dimensional surface instead of the whole volume. Recently an effective numerical method based on the BEM was developed for the fast simulation of normal and tangential contacts including indentation test and partial sliding of fractal rough surfaces (Pohrt & Li, 2014). This method was further developed for the adhesive contact by formulating a mesh-dependent stress criterion for detachment of contact elements (Pohrt & Popov, 2015). In the present paper, we develop the method suggested by Pohrt & Popov for the contact of functionally graded materials with power-law dependence of the elastic modulus. To our knowledge, the presently proposed method is the first formulation of adhesive BEM for graded materials allowing simulating contacts of arbitrary shape. Already existing studies on adhesion of FGMs are restricted solely to analytical analysis of single axis-symmetric contacts or ensembles of such contacts as the Greenwood–Williamson like models (Heß, 2016; Jin & Guo, 2013; Jin, Guo & Zhang, 2013; Jin, Zhang, Wan & Guo, 2016).

# 2. Boundary Element Method for normal contact of power-law graded materials

## 2.1 Fundamental solution and calculation of influence matrix

Let us consider a nonhomogeneous elastic half space, whose elastic modulus depends on the depth $z$ according to a power law:

$$E(z) = E_0 \left(\frac{z}{c_0}\right)^k, \text{ with } -1 < k < 1, \tag{1}$$

where $c_0$ is a characteristic length and $E_0$ a constant having the unity of elastic modulus. The case with the power $k = 0$ corresponds to the homogeneous elastic material. If $0 < k < 1$, the elastic modulus is increasing with depth and for $-1 < k < 0$ it is decreasing with depth.

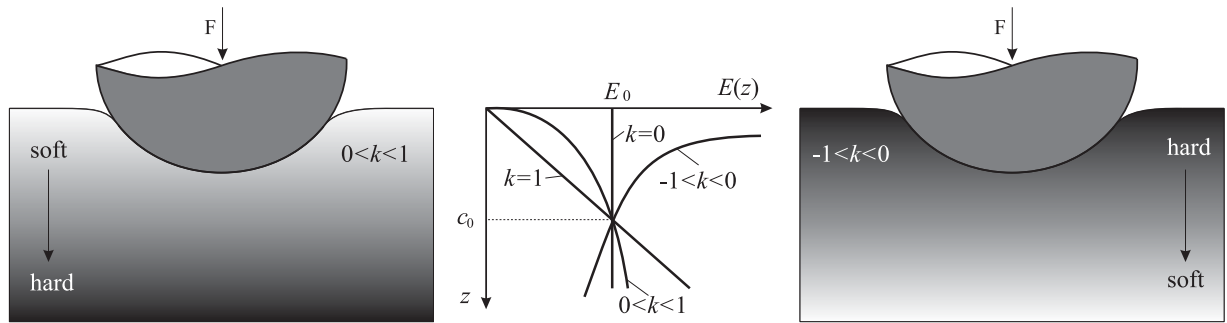

**Figure 1 Contact between a rigid body and a power law elastic half space.**

The fundamental solution for a power-law gradient medium was given by Booker et al. (Booker, Balaam & Davis, 1985a) and Giannakopoulos et al. (Giannakopoulos & Suresh, 1997a): the vertical surface deflection of this elastic half space $u_z(x, y)$ at a distance $r$ from the the point $(x', y')$ of action of a concentrated force $F_N$ is given by

$$u_z(x, y) = \frac{\alpha(k, \nu) \cdot (1 - \nu^2) \cdot c_0^k}{E_0} \cdot \frac{F_N}{r^{1+k}}, \tag{2}$$

where $\nu$ is Poisson's ratio, $E_0$ elastic modulus, and $r = \sqrt{(x-x')^2 + (y-y')^2}$. $\alpha$ is a function of the power $k$ and Poisson's ratio $\nu$ (which is assumed to be constant):

$$\alpha(k, \nu) = \frac{2^k \beta \sin\left(\frac{\beta\pi}{2}\right) \Gamma\left(\frac{3+k+\beta}{2}\right) \Gamma\left(\frac{3+k-\beta}{2}\right)}{\sqrt{\pi}(1+k)\cos\left(\frac{k\pi}{2}\right) \Gamma(2+k) \Gamma\left(1+\frac{k}{2}\right) \Gamma\left(\frac{1-k}{2}\right)} \tag{3}$$

with

$$\beta(k, \nu) = \sqrt{(1+k)\left(1 - \frac{k\nu}{1-\nu}\right)} \tag{4}$$

and Γ the gamma function. Note that equation (2) is valid for any power in the range of $-1 < k < 1$. However, in the most previous studies only positive powers in a range of $0 \le k < 1$ were considered. The reason for that is possibly the fact that in the soil contact problems –the starting point of these studies – the elastic modulus of the ground increases with the soil depth. The researchers in soil mechanics and geomechanics thus have mainly paid attention to modeling the cases where elastic modulus is increasing with depth up to the limiting case of $k = 1$ ("Gibson soil model") when the elastic modulus increases linearly with depth (Burland, Longworth & Moore, 1977). Lee and Baber et al. suggested to remove the restriction of extend the solution of Giannakopoulos et al. (Giannakopoulos & Suresh, 1997a) to $-1 < k < 1$ (Lee, Barber & Thouless, 2009).

For any distributed normal load $p(x', y')$ acting in an area $A$, the vertical displacement $u_z(x, y)$ can be calculated by using the superposition principle:

$$u_z(x,y) = \frac{\alpha(k,\nu)(1-\nu^2)c_0^k}{E_0} \cdot \iint_A \frac{p(x',y')}{r^{1+k}} dx' dy' . \qquad (5)$$

In the Boundary Element Method, the integral (5) is estimated by discretizing the integration area. Let us divide the surface in rectangular elements having the size $\Delta x$ and $\Delta y$ in the direction of coordinate axes (Figure 2). In the discrete calculation, we assume the the pressure inside each element is constant. For the homogeneous medium, the displacement of the surface under the action of a uniform pressure in a rectangular area was first given by Love in 1929 (Love, 1929), and can also be found in the book by Johnson (Johnson, 1987). Now we repeat the same calculation for the nonhomogeneous case.

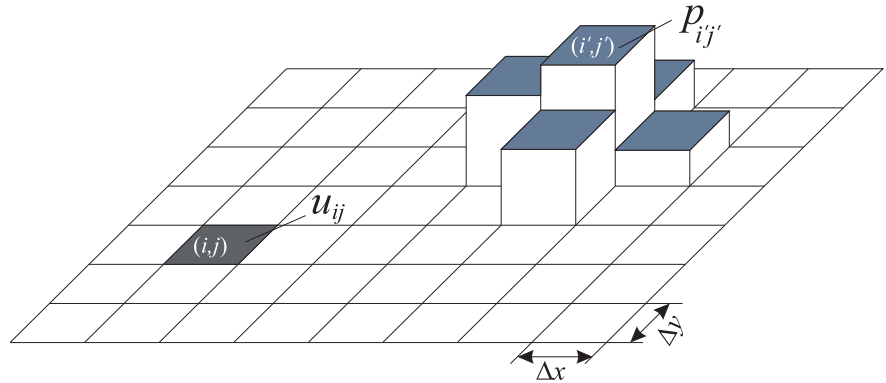

**Figure 2 Discretization of simulation area into rectangular elements loaded under the uniform pressures.**

Substituting constant pressure into Eq. (5), we obtain the vertical displacement $u_{i,j}$ of the surface at the position $(x_i, y_j)$ under the uniform pressure $p_{i',j'}$ acting in the unit cell at the position $(x_{i'}, y_{j'})$:

$$u_{i,j} = \frac{\alpha(k,\nu)(1-\nu^2)c_0^k}{E_0} \cdot \int_{y_{j'}-\Delta x/2}^{y_{j'}+\Delta x/2} \int_{x_{i'}-\Delta x/2}^{x_{i'}+\Delta x/2} \frac{1}{\left[(x_i - x_{i'}) + (y_i - y_{j'})\right]^{\frac{1+k}{2}}} dx_{i'} dy_{j'} \cdot p_{i',j'} . \qquad (6)$$

For an arbitrary discretized pressure distribution, we can write

$$u_{i,j} = \frac{\alpha(k,v)(1-v^2)c_0^k}{E_0} \cdot \sum_{i'}\sum_{j'} K_{i,j,i',j'} p_{i',j'} \qquad (7)$$

with the coefficient

$$K_{i,j,i',j'} = \int_{y_{j'}-\Delta x/2}^{y_{j'}+\Delta x/2} \int_{x_{i'}-\Delta x/2}^{x_{i'}+\Delta x/2} \frac{1}{\left[(x_i - x_{i'})+(y_i - y_{i'})\right]^{\frac{1+k}{2}}} dx_{i'} dy_{j'} . \qquad (8)$$

By changing the integration variables, we come to the expression

$$K_{i,j,i',j'} = \int_n^m \int_q^p \frac{1}{(a^2+b^2)^{\frac{1+k}{2}}} da\,db \qquad (9)$$

with $p=(i'-i+0.5)\cdot\Delta x$, $q=(i'-i-0.5)\cdot\Delta x$, $m=(j'-j+0.5)\cdot\Delta y$, $n=(j'-j-0.5)\cdot\Delta y$. It can be seen that the coefficient is the function of element size, distance between positions of the wanted displacement and the load, and the power $k$:

$$K_{i,j,i',j'} = K_{i,j,i',j'}(i-i', j-j', \Delta x, \Delta y, k). \qquad (10)$$

This "influence matrix" has to be calculated only once during any calculation. Integration of Eq.(9) gives

$$K = \frac{1}{mn(k-k^2)} \cdot$$
$$\left[ -np(m^2+p^2)^{\frac{1-k}{2}} \cdot {}_2F_1\left(\frac{1}{2},1;\frac{k+2}{2};-\frac{p^2}{m^2}\right) + mp(n^2+p^2)^{\frac{1-k}{2}} \cdot {}_2F_1\left(\frac{1}{2},1;\frac{k+2}{2};-\frac{p^2}{n^2}\right) \right.$$
$$+ nq(m^2+q^2)^{\frac{1-k}{2}} \cdot {}_2F_1\left(\frac{1}{2},1;\frac{k+2}{2};-\frac{q^2}{m^2}\right) - mq(n^2+q^2)^{\frac{1-k}{2}} \cdot {}_2F_1\left(\frac{1}{2},1;\frac{k+2}{2};-\frac{q^2}{n^2}\right) \qquad (11)$$
$$+ knp(m^2+p^2)^{\frac{1-k}{2}} \cdot {}_2F_1\left(1,1-\frac{k}{2};\frac{3}{2};-\frac{p^2}{m^2}\right) - kmp(n^2+p^2)^{\frac{1-k}{2}} \cdot {}_2F_1\left(1,1-\frac{k}{2};\frac{3}{2};-\frac{p^2}{n^2}\right)$$
$$\left. - knq(m^2+q^2)^{\frac{1-k}{2}} \cdot {}_2F_1\left(1,1-\frac{k}{2};\frac{3}{2};-\frac{q^2}{m^2}\right) + kmq(n^2+q^2)^{\frac{1-k}{2}} \cdot {}_2F_1\left(1,1-\frac{k}{2};\frac{3}{2};-\frac{q^2}{n^2}\right) \right]$$

where ${}_2F_1(a,b;c;z)$ is the Gaussian hypergeometric function. In the simulation, we found that the evaluation of the hypergeometric functions in Eq. (11) takes more time than the direct numerical integration of (9).

Let us write Eq. (7) in the matrix form

$$\mathbf{u} = \frac{\alpha(k,\nu)(1-\nu^2)c_0^k}{E_0} \cdot \mathbf{Kp}. \tag{12}$$

If the size of the calculation is $N \times N$, then the matrix $\mathbf{K}$ has the size $N^4$ and the complexity of the problem will be on the order $\mathrm{O}(N^4)$. To reduce the computation costs, two mathematical methods have been suggested: the multigrid method, which is e.g. used for dry contact and elastohydrodynamic lubrication by Venner and Lubrecht (Venner & Lubrecht, 2000), and the Fast Fourier Transform (FFT), e.g. applied for rough contact (Stanley & Kato, 1997). Both methods have the complexity $\mathrm{O}(N^2 \log N)$. Similarly to (Pohrt & Li, 2014), we use here the latter technique due to the easier programing and the possibility to carry out parallel calculation on a graphics processing unit. Using the FFT, Eq. (7) is calculated as

$$\mathbf{u} = \frac{\alpha(k,\nu)c_0^k}{E_0} \cdot \mathrm{IFFT}\left[\mathrm{FFT}(\mathbf{K}) \cdot \mathrm{FFT}(\mathbf{p})\right]. \tag{13}$$

The inverse problem, finding the pressure necessary to generate the given surface deformation, can be theoretically solved from the calculation of the inverse matrix in Eq.(12). However, the method which is used practically is the so-called conjugate-gradient (CG) method, which was introduced by Polonsky and Keer for normal rough contact (Polonsky & Keer, 1999) and recently applied by Pohrt and Li also for tangential contacts including the partial sliding of rough bodies (Pohrt & Li, 2014). The detailed algorithm of the conjugate-gradient method can be found in the paper (Pohrt & Li, 2014).

## 2.2 Examples of numerical simulations

First, we start with the 'Hertzian' contact problem with a power law elastic half space. A parabolic indenter is pressed into the elastic half space with a normal force $F_N$, which results in an indentation depth $d$ and a contact radius $a$. The analytical solution for surface displacement, pressure distribution, as well as for relations between $F_N$, $d$ and $a$ can be found in the paper (Heß, 2016). For convenience of the reader and further reference, we reproduce here this solution: The surface deflection of elastic half space inside and outside of the contact area is given by

$$\frac{u_z(r,a)}{a^2/R} = \begin{cases} \dfrac{1}{k+1} - \dfrac{r^2}{2a^2}, & r < a \\ \dfrac{\cos(k\pi/2)}{\pi(k+1)}\left[\mathrm{B}\left(\dfrac{a^2}{r^2}, \dfrac{1+k}{2}, \dfrac{1-k}{2}\right) - \dfrac{r^2}{a^2}\mathrm{B}\left(\dfrac{a^2}{r^2}, \dfrac{3+k}{2}, \dfrac{1-k}{2}\right)\right], & r > a \end{cases} \tag{14}$$

The dependences of the indentation depth and normal the load on contact radius are

$$d(a) = \frac{1}{k+1} \cdot \frac{a^2}{R}, \tag{15}$$

$$F_N(a) = \frac{4\alpha(k,\nu)}{\Gamma\left(\frac{1+k}{2}\right)\Gamma\left(\frac{1-k}{2}\right)(k+1)^2(k+3)} \frac{E_0}{(1-\nu^2)c_0^k} \cdot \frac{a^{3+k}}{R}. \qquad (16)$$

Now we reproduce this solution using the above described BEM. The simulation area is divided into 512x512 elements. For each given indentation depth, the pressure distribution, surface deformation of half space as well as the contact area and the normal load are calculated. Figure 3(a) shows the displacement of the elastic half space for media with different power $k$. To clearly see the consistency, the numerical results are shown in the figure only at a few surface points. Figure 3(b) shows the dependency of the normal force on the indentation depth for elastic half spaces with powers $k$ from -0.8 to 0.8. Note that the following dimensionless variables are used in Figure 3(b):

$$\tilde{d} = d/R, \tilde{F}_N = \frac{\Gamma\left(\frac{1+k}{2}\right)\Gamma\left(\frac{1-k}{2}\right)(k+1)^2(k+3)}{4\alpha(k,\nu)(1+k)^{\frac{3+k}{2}}R^{2+k}} \frac{(1-\nu^2)c_0^k}{E_0} F_N. \qquad (17)$$

From Figure 3 we can see that numerical results agree with the existing analytical solutions very well. With this method we can simulate the contact of rigid bodies having complicated geometries, such as indenters with polygonal cross-sections or fractal surfaces. An example is seen in Figure 4 with 1024x1024 meshing grids. The contact areas between a rough surface and different power law elastic half spaces are obtained at the same indentation depth. One advantage of this method is the fast computing using the techniques described in section 2.1. For example, one single indentation step in Figure 3b takes only about one second.

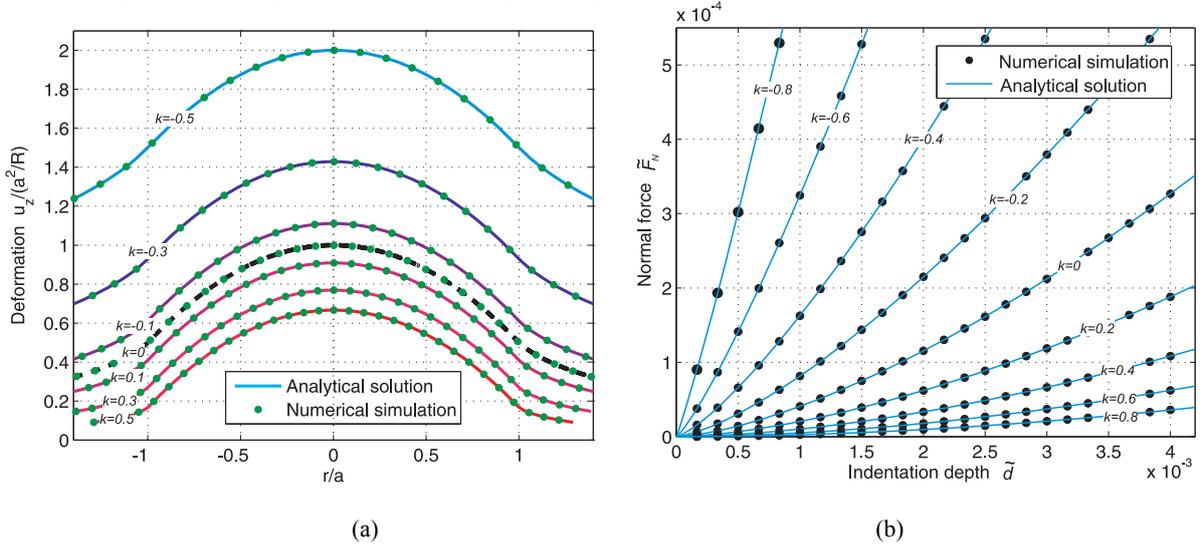

(a)          (b)

**Figure 3 Contact between a rigid sphere and an elastic half space with different powers: (a) surface deformation; (b) dependence of normal load on indentation depth.**

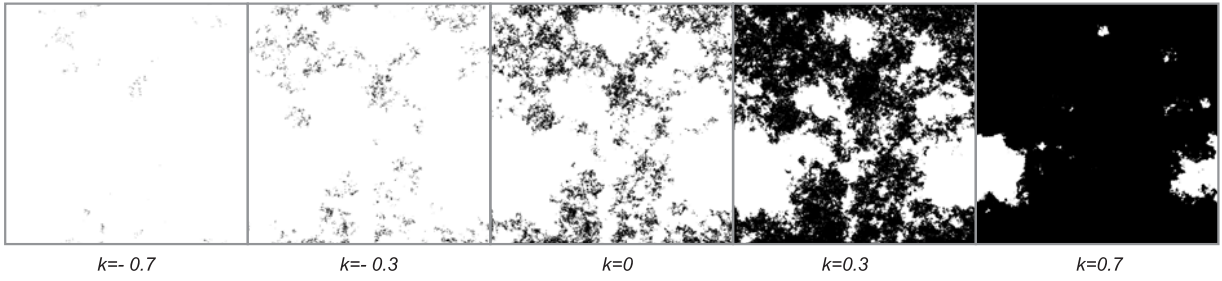

| k=-0.7 | k=-0.3 | k=0 | k=0.3 | k=0.7 |

**Figure 4** Contact areas of a rough surface pressed into elastic half spaces at the same indentation depth.

# 3. Adhesive contact of the functionally graded materials

## 3.1 Mesh dependent detachment criterion and adhesive BEM

Recently a local detachment criterion was proposed by Pohrt and Popov for the numerical simulation of adhesive contact with the boundary element method (Pohrt & Popov, 2015). The idea is based on the Griffith's energy balance: the crack (or in our case the boundary of the adhesive contact) is in the equilibrium if the released elastic strain energy is equal to the energy required to generate two new surfaces (work of adhesion). In the BEM, we consider the rectangular elements in contact. The element will detach if the energy released by its detaching exceeds the work of adhesion $U_{el} > U_{surf}$. The released elastic energy can be easily calculated

$$U_{el} = \frac{1}{2}\iint_A p u_z \mathrm{d}A, \tag{18}$$

where $p$ is the pressure on the element, $u_z$ the displacement caused by this pressure. In the case of graded materials, the deformation is given by Eq.(5), and pressure is assumed to be constant on a single grid. Then the Eq. (18) can be written as

$$U_{el} = \frac{\alpha(k,\nu)(1-\nu^2)c_0^k p^2}{2E_0}\int_0^{\Delta y}\int_0^{\Delta x}\int_0^{\Delta y}\int_0^{\Delta x}\frac{1}{r^{1+k}}\mathrm{d}x\mathrm{d}y\mathrm{d}x'\mathrm{d}y'. \tag{19}$$

The work of adhesion is simply equal to

$$U_{surf} = \gamma_{12} \cdot \Delta x \Delta y, \tag{20}$$

where $\gamma_{12}$ is the relative surface energy (work of adhesion per unit area).

For square grid elements $\Delta x = \Delta y = \Delta$, Eqs. (19) (20) simplifies to

$$U_{el} = \frac{\alpha(k,\nu)(1-\nu^2)c_0^k p}{2E_0} \cdot \Delta^{3-k} \cdot c_1, \tag{21}$$

$$U_{surf} = \gamma_{12} \cdot \Delta^2, \tag{22}$$

with

$$c_1(k) = \int_0^1 \int_0^1 \int_0^1 \int_0^1 \frac{1}{\left[(a-a')+(b-b')\right]^{\frac{1+k}{2}}} da\, db\, da'\, db', \tag{23}$$

here the function $c_1(k)$ is a dimensionless coefficient depending only on the power $k$. Results of numerical integration of $c_1(k)$ are shown in Figure 5 (dashed line). Comparing the elastic energy and surface energy, $U_{el} = U_{surf}$, we can obtain a critical stress

$$p_{crit} = \sqrt{\frac{2E_0 \gamma_{12}}{\alpha(k,\nu)(1-\nu^2)c_1 c_0^k \Delta^{1-k}}}. \tag{24}$$

In the simulation, if the pressure on the element in contact is larger than this critical value, $p > p_{crit}$, the element separates from the contact.

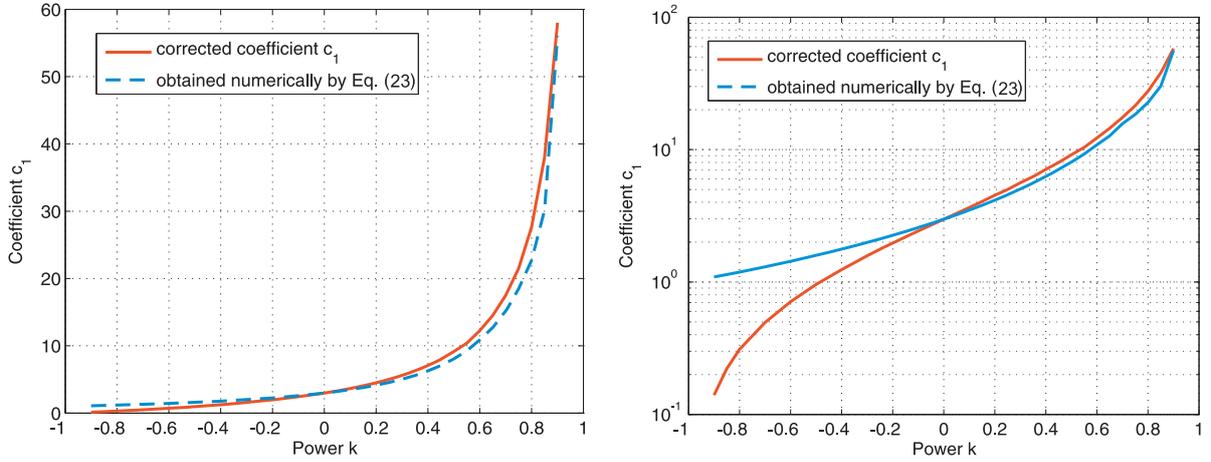

**Figure 5 Coefficient $c_1$ for different power $k$. The difference from the corrected coefficient can be clearly seen in the logarithmic coordinate in right figure.**

The algorithm for a simulation of adhesive contact is now the following. The indenter is initially pressed into the elastic half space to depth $d_0$ without consideration of adhesion, which results in surface displacement $u_{z0}(i,j)$ and some contact area $A_0$. This non-adhesive normal contact problem can be solved as described in section 2. Now we pull off the indenter by some incremental depth $\Delta d$. In each calculation step, the contact area is first considered unchanged, and the pressure $p$ corresponding to the assumed incremental indentation depth: $u_{z0}(i,j) - \Delta d$ is calculated in all points of the calculation net in the contact area $A_0$ as well as displacements $u_{z0}(i,j)$ outside the contact area (this part of the contact problem can be solved by the conjugate-gradient method). Then the criterion (24) is checked: all elements satisfying condition $p > p_{crit}$ are separated from the contact, and a new contact area $A'$ is obtained. With this new area we come back to the inverse contact problem, and recalculate the pressure $p'$ in the contact area. The procedure is repeated until the pressures in all elements meet the condition $p \leq p_{crit}$. Then the normal force can be easily calculated by sum of pressures, and the simulation comes to the next pull-off step.

By comparing to the existing analytical solutions, we found that the exact coincidence between numerical and analytical results is achieved if the $c_1$ in (24) is corrected. This may be due to the fact that the coefficient (23) is as a matter of fact determined not unambiguously in a discrete system. It is not clear if the displacement in the integral has to be taken in the center of the calculation cell or it should be taken some other value. However, we found that there always exists some $c_1(k)$ providing exact coincidence with analytical solution. We will call this value of $c_1(k)$ the "corrected $c_1(k)$". It is shown in Figure 5 with solid line and listed in Table 1. With the corrected coefficient $c_1(k)$, the simulations for various changes in parameters such as material parameters, surface energy, simulation area, geometry parameters, give a good agreement with theoretical solutions.

Table 1 Values of corrected coefficient $c_1$

| power $k$ | corrected $c_1$ | power $k$ | corrected $c_1$ | power $k$ | corrected $c_1$ |
|---|---|---|---|---|---|
| 0.9 | 58.00 | 0.2 | 4.52 | -0.5 | 0.96 |
| 0.8 | 27.65 | 0.1 | 3.66 | -0.6 | 0.71 |
| 0.7 | 17.50 | 0 | 2.97 | -0.7 | 0.495 |
| 0.6 | 12.25 | -0.1 | 2.43 | -0.8 | 0.31 |
| 0.5 | 9.15 | -0.2 | 1.97 | -0.9 | 0.14 |
| 0.4 | 7.10 | -0.3 | 1.58 | | |
| 0.3 | 5.62 | -0.4 | 1.24 | | |

## 3.2 Examples of numerical simulations

Firstly we simulate the adhesive contact of flat cylindrical punch with the radius $a^*$ and an elastic half space. The contact radius during the pull-off process keeps unchanged, $a = a^*$. The analytical solution of the critical distance of separation and the critical normal force is given in the paper (Heß, 2016)

$$d_{crit} = \sqrt{\alpha(k,\nu)\Gamma\left(\frac{1-k}{2}\right)\Gamma\left(\frac{1+k}{2}\right)\frac{2\pi(1-\nu^2)\gamma_{12}a^{1-k}c_0^k}{E_0}}, \qquad (25)$$

$$F_{crit} = \sqrt{\alpha(k,\nu)\Gamma\left(\frac{1-k}{2}\right)\Gamma\left(\frac{1+k}{2}\right)\frac{8\pi\gamma_{12}E_0}{(1-\nu^2)(k+1)^2 c_0^k}} \cdot a^{\frac{3+k}{2}}. \qquad (26)$$

Figure 6 shows an example of indentation controlled contact for graded elastic half space with $k = -0.5$. The change of the normal force and the contact radius during the process is presented in Figure 6b, in which the direction of arrows shows the pull-off process, the red triangles indicate the contact state specified in Figure 6a. Note that the normalized coordinates are used in Figure 6b so that the one can compare the simulation with theoretical solutions (25) and (26).

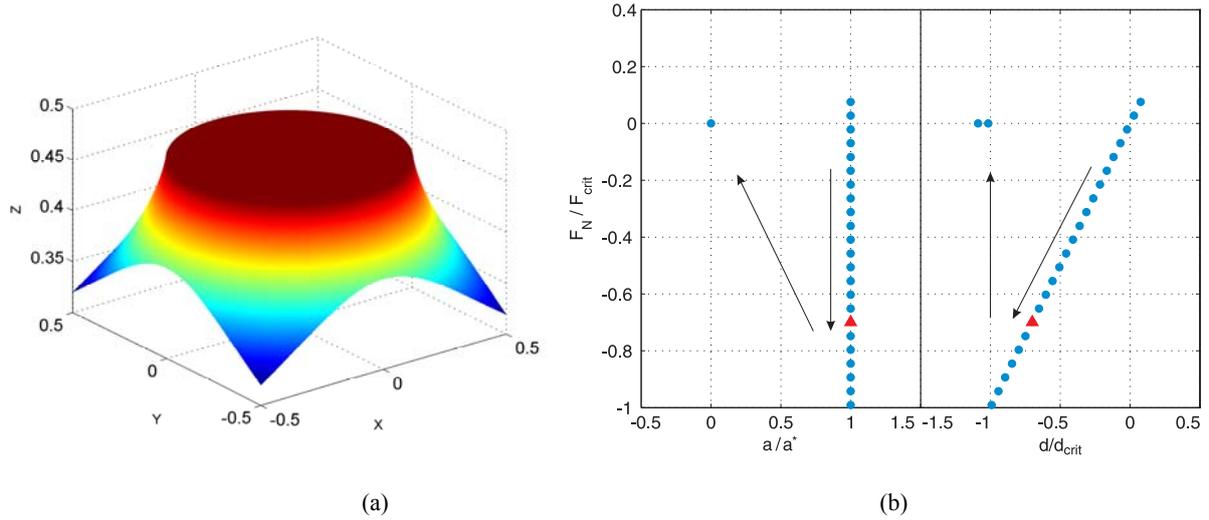

(a)             (b)

**Figure 6 Adhesive pull-off contact between a rigid flat ended cylinder and elastic half space with $k$=-0.5: (a) stretching of elastic half space; (b) change of normal force and contact radius during the pull-off indentation.**

In Figure 7, results of simulation of indentation of a parabolic indenter into a gradient medium with the power $k = 0.5$ are shown. Different meshing grids with 256x256 and 1024x1024 are used for this example. The contact radius $a$ as well as normal force $F_N$ as the function of indentation depth during the pull-off process is shown in Figure 7. The adhesive behavior for different values of $R/c_0$ is shown in Figure 8a, and for different powers $k$ of elastic half space in Figure 8b with a meshing grid 512x512. The corresponding analytical solutions can be found in the papers (Heß, 2016; Jin, Guo & Zhang, 2013). It can be seen that the numerical simulation agrees with the theoretical solution very well.

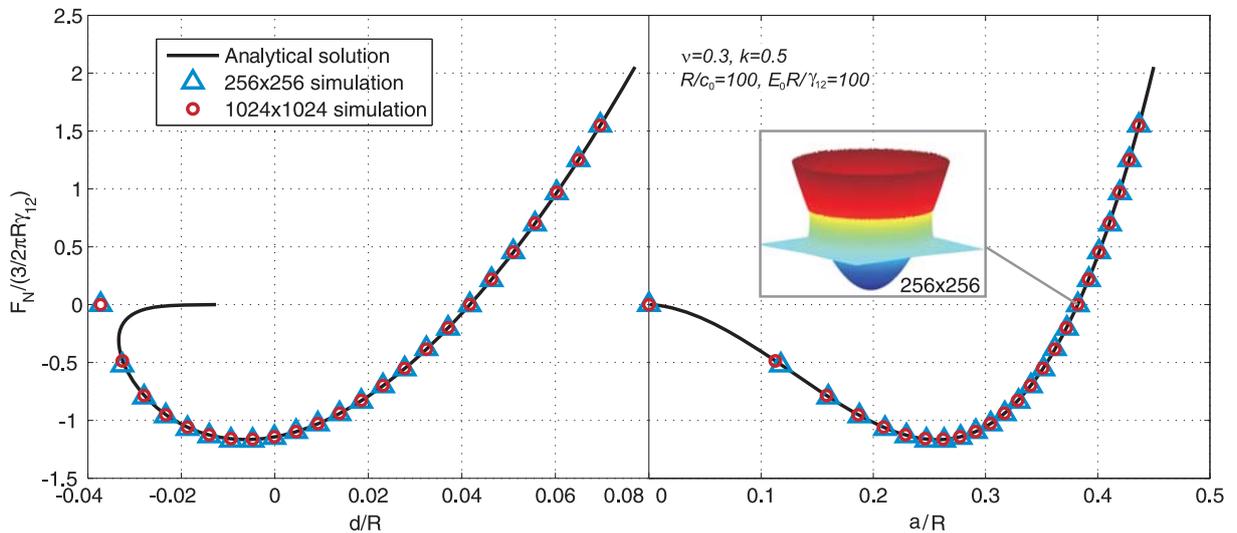

**Figure 7 Normal force and contact radius during the pull-off adhesive indentation between a rigid parabolic indenter and elastic half space with 256x256 and 1024x1024 meshing grids.**

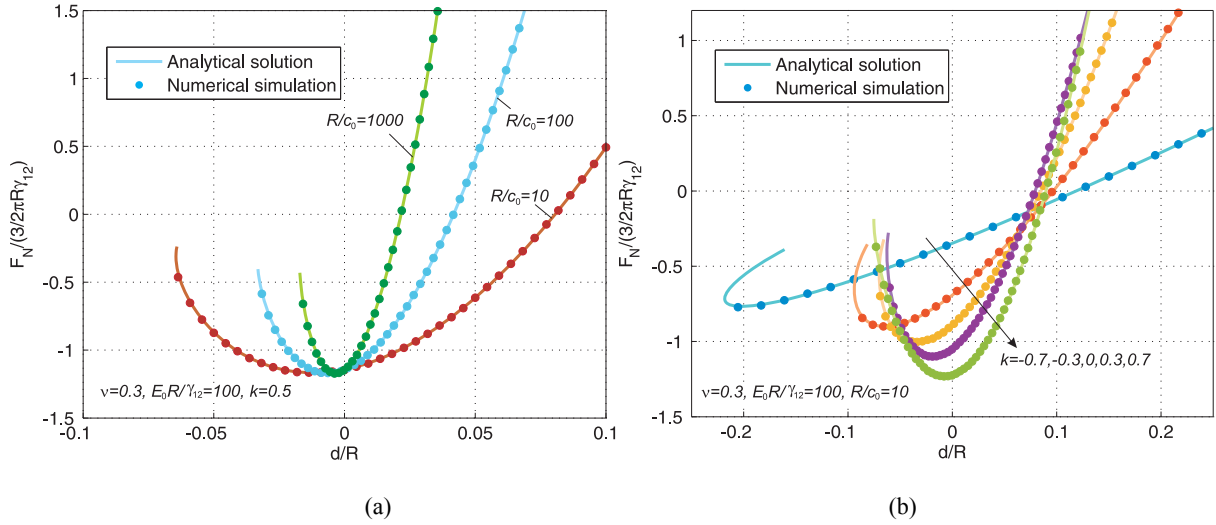

(a)                                (b)

**Figure 8** Dependence of normal force on the indentation depth for the contact between a parabolic indenter and different graded elastic half spaces (512x512 meshing grid).

Finally we would like to present some test showing the independence of the macroscopic results on the size and orientation of the simulation net. Although the detachment criterion (24) in adhesive contact is meshing size dependent, the size of the unit cell may not have influence on the macroscopic behavior of the system. This independence is illustrated in Figure 7 where simulation results with meshing grids 256x256 and 1024x1024 are presented in the same plot. They coincide exactly. Further, due to discretization there could be some effects of "pinning" the contact area to the grid. The absence of such effects is supported by the fact that the numerically obtained contact areas for axis-symmetric indenters are axis-symmetric too Figure 9. Another test of the independence of numerical issues is presented in Figure 10 showing results of simulation of an adhesive contact between a square punch (gray color) and a graded elastic half space with power $k$=0.3 (this simulation was carried out with the grid 512x512). The orientation of meshing grid (see the small square elements) and placement of square indenter are shown in Figure 10a. We simulate the adhesive pull-off indentation of the punch with four different angles between the orientation of the square and the grid lines (from 0 to 45°). Contact areas (black color) at the moment of final detachment are shown in Figure 10a, and the values of normal force and contact areas during the pull-off in Figure 10b and Figure 10c. The configuration of the final state does not depend on the orientation between the square and the grid lines. The normal force and indentation depth in Figure 10b and Figure 10c are normalized to the critical values for corresponding inscribed cylindrical punch. The adhesive behavior of indenter with a square cross-section can then be compared with a cylindrical punch having the same area.

Using the proposed method, the adhesive contact between indenters with various shapes and elastically graded half space can be simulated, for example the power-law shape of axisymmetric indenters, which for homogeneous media was firstly analyzed by Borodich and Galanov (see Borodich, Galanov & Suarez-Alvareza, 2014), and for toroidal indenters in paper (Argatov, Li, Pohrt & Popov, 2016) or rough surfaces. It is in particular fast for simulation of flat 'stamps' with different geometries, because the pull-off process can begin with zero indentation depth.

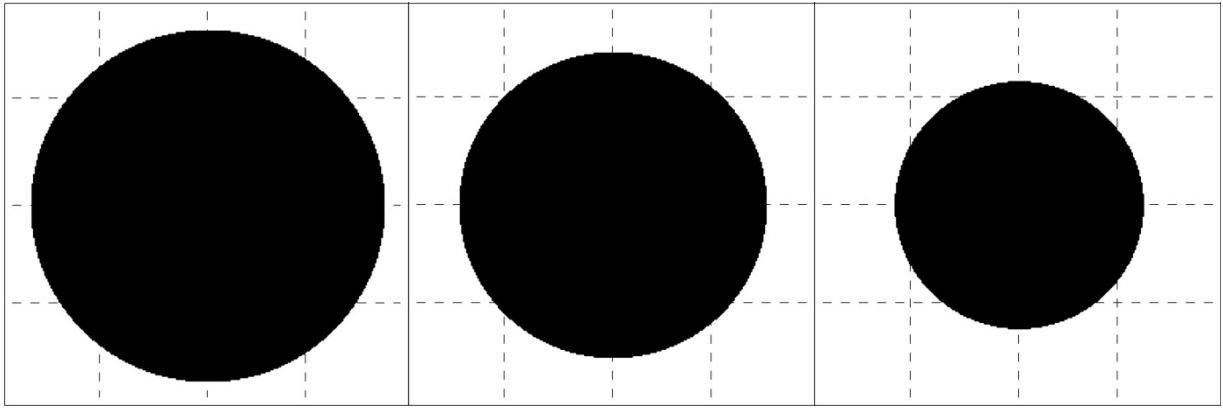

**Figure 9 Three consecutive contact areas in a 256x256 simulation of contact with a parabolic indenter.**

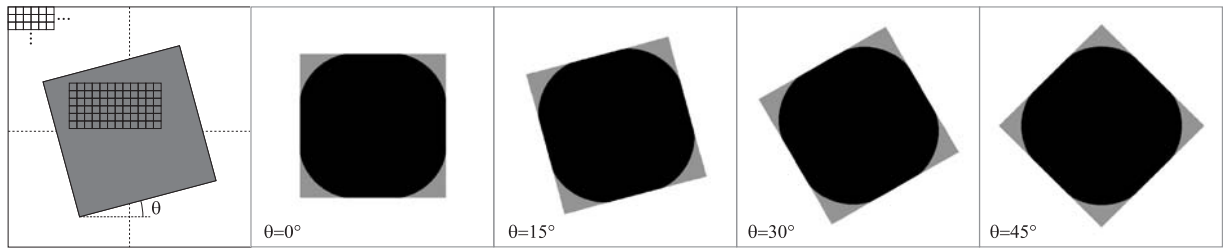

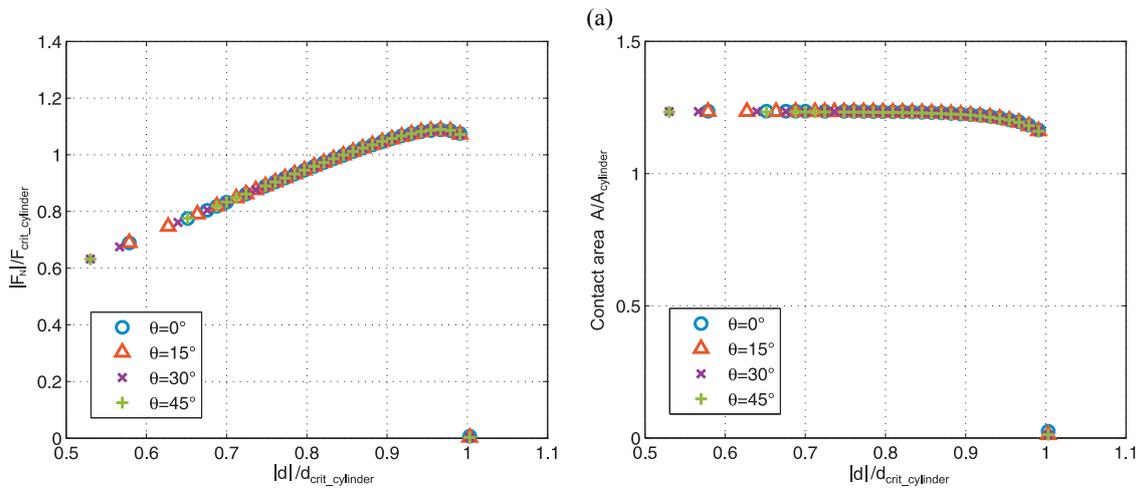

**Figure 10 512x512 simulation of adhesive pull-off indentation of a square punch in different orientations with a graded elastic half space with *k*=0.3: (a) contact areas at the detachment moment; (b) normal force and (c) contact area during the pull-off process in normalized coordinate.**

## 4. Conclusion

We developed the adhesive BEM previously proposed by Pohrt and Popov for the case of contact with power-law graded materials. In particular, the influence matrix was calculated, and a local mesh-size dependent detachment criterion was proposed for the adhesive contact. The calculation of displacement through the given stress distribution was implemented using the FFT, which enables a drastic acceleration of numerical calculation. For verifying the proposed method, several exactly analytically solved problems were reproduced numerically.

The simulation results have been shown to be independent on the mesh size and the orientation of the numerical grid.